\documentclass[a4paper,11pt]{article}
\usepackage[utf8]{inputenc}
\usepackage{pos}
\usepackage{slashed, physics, mathtools, xfrac, nicefrac, tabu}
\usepackage{textpos, svg, subcaption, tikz, booktabs, dsfont}
\usetikzlibrary{arrows, shapes}
\usetikzlibrary{positioning}
\usetikzlibrary{decorations.markings}
\usetikzlibrary{patterns,math,calc}
\usetikzlibrary{decorations.pathreplacing}
\usetikzlibrary{positioning}
\usetikzlibrary{decorations.pathmorphing}
\usepackage[noabbrev]{cleveref}
\usepackage{xcolor}

\newcommand{\comment}[1]{}
\newcommand{\me}{\textrm{e}}
\newcommand{\inter}{\mathcal{N}}

\title{Nucleon Charges and Sigma Terms from \(N_{f}=2+1\) QCD}

\author*{Daniel Jenkins}
\author{Sara Collins}
\author{Gunnar Bali}

\affiliation{Institut f\"ur Theoretische Physik,\\
  Universit\"at Regensburg,\\
  Universit\"atsstraße 31, 93053, Regensburg, Germany}

\emailAdd{daniel.jenkins@ur.de}
\emailAdd{sara.collins@ur.de}
\emailAdd{gunnar.bali@ur.de}

\abstract{We report on recent progress of our analysis of the nucleon sigma terms, as well as the singlet scalar, axial and tensor nucleon charges.
  These are determined employing the CLS gauge ensembles, which are generated using the L\"uscher-Weisz gluon action and the non-perturbatively improved Sheikholeslami-Wohlert fermion action with \(N_{f}=2+1\) dynamical fermions.
  For the ensembles analysed thus far, the pion masses range from 200 MeV up to 410 MeV, and the lattice spacings take five values between 0.09~fm and 0.04~fm.
  We have employed a variety of methods to determine the relevant correlation functions, including the sequential source method for connected contributions and the truncated solver method for disconnected contributions.
}

\FullConference{%
 The 38th International Symposium on Lattice Field Theory, LATTICE2021
  26th-30th July, 2021
  Zoom/Gather@Massachusetts Institute of Technology
}

\begin{document}
\maketitle

\section{Introduction}\label{sec:intro}
The nucleon charges~($g^q_X$) are matrix elements of the form \(\expval{J}{N}\), where the current $J=\overline{q}\Gamma q$ is composed of spinor fields of quark flavour \(q \in \{u,d,s\}\) and  \(\Gamma \in \{ \mathds{1}, \gamma^{5}\gamma^{\mu},\gamma^{\mu},\frac{i}{2}[\gamma^{\mu}\gamma^{\nu}] \}\) for the scalar, axial, vector, and tensor charges~($X=S, A, V$ and $T$), respectively.  Of particular note are the sigma terms which are obtained from
the scalar charges through the multiplication with the mass of the quark $m_q$:  \(\sigma_q=m_{q}g_S^q=m_q\expval{\overline{q}\mathds{1}q}{N}\). These are of interest, for example, as they appear in the decomposition of the nucleon mass~\cite{Ji_1995}~(representing the quark contribution to the mass), and are required to
predict the spin-independent WIMP-nucleon scattering cross section relevant for dark matter detection experiments. The axial charges give the contributions of the quark spins to the spin of the nucleon~(as well as the coupling to the $Z$-boson), while the tensor charges correspond to the quark transverse spins in the nucleon.

\section{Lattice Setup}\label{sec:lattice}
We utilise the \(N_{f}=2+1\) CLS ensembles~\cite{Bruno_2015} within
our analysis, which were generated with the Lüscher-Weisz gluonic
action, and the non-perturbatively improved Sheikholeslami-Wohlert
fermionic action. There are several important aspects of this setup.
We have \(\mathcal{O}(a)\) non-perturbative improvement of the fermion
action, however, the currents also require improvement, which, in
general, involves both quark mass dependent and independent terms. As
we are working in the forward limit, the latter do not appear (as they
involve derivatives). The only exception to this is the scalar
current, for which the mass-independent improvement term is
proportional to \(aFF\). We omit this term as the corresponding
coefficient has not yet been determined. For this
preliminary analysis, the mass-dependent terms are also not
considered~(for all of the charges).  Note that, due to chiral
symmetry breaking, there can be mixing between quark flavours under
renormalisation. This is discussed in \cref{sec:renorm}.

\begin{figure}[ht]
  \centering
  \includegraphics[width=0.7\textwidth]{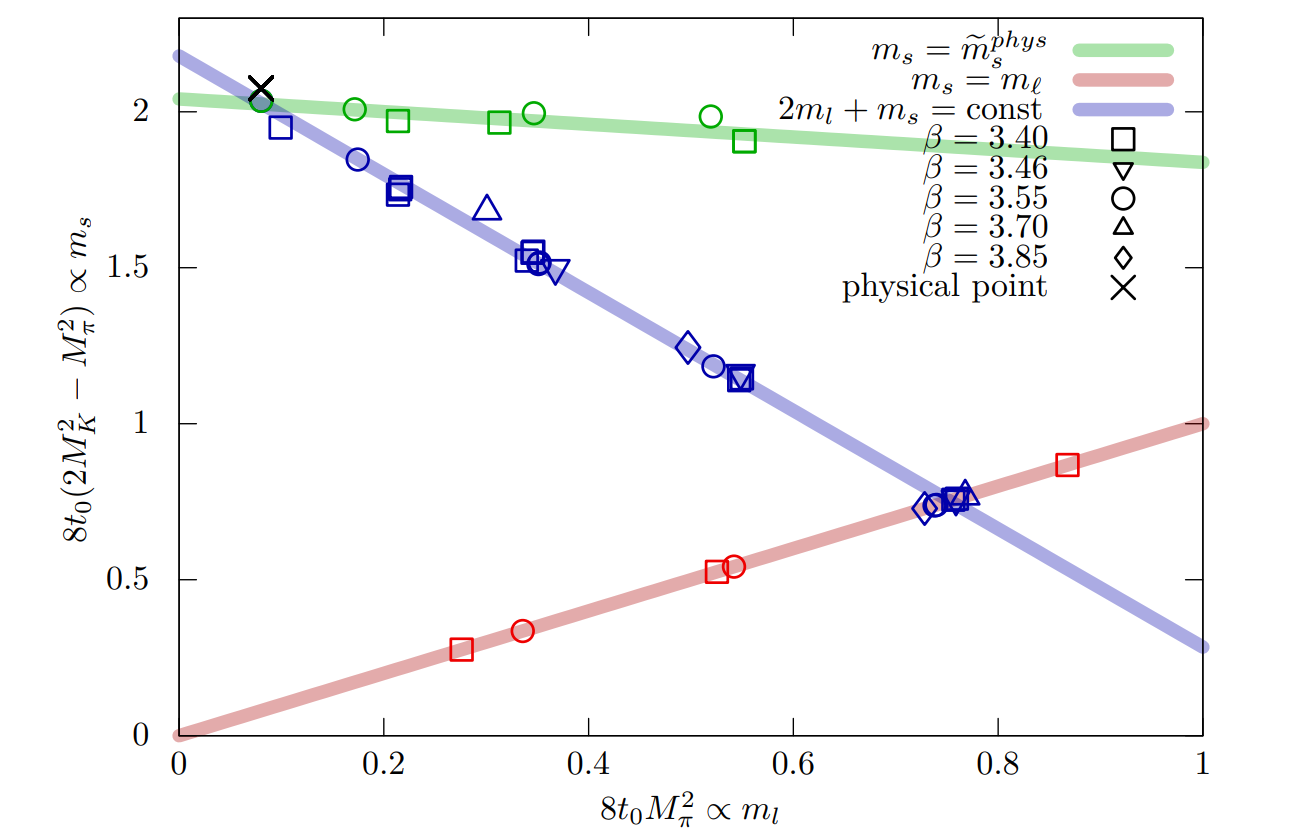}
  \caption{\label{fig:ens} Overview of the CLS ensembles utilised within this analysis.}
\end{figure}

The light and strange quark masses are varied in the simulation so as
to follow three trajectories. As shown in \cref{fig:ens}, two of the
trajectories approach the physical point~(one along which the flavour
average quark mass is held constant and the other along which the
physical strange quark mass is kept approximately constant), while the
other approaches the chiral limit. This allows for full control of
quark-mass systematics.  Furthermore, the range of lattice spacings
(spanning 0.09~fm down to 0.04~fm) and volumes (with \(L^3\cdot T=24^{3}\cdot
48\) up to \(96^{3}\cdot 192\), with \(Lm_{\pi} \gtrsim 4\) in almost all
cases) available enables discretisation and finite volume effects to
be thoroughly investigated. A high statistics study can be realised,
with each ensemble typically containing around 1000--2000
configurations. Note that, in order to counter-act topological
freezing, many ensembles (in particular, those at finer lattice
spacing) have open boundary conditions in the time direction.

\section{Correlation Functions}\label{sec:contri}
The correlation functions used are the standard two- and three-point functions
\begin{align}
  &C_{2pt}(t_f, t_i) = \expval{\inter(t_f) \overline{\inter}(t_i)}, \label{equ:c2}\\
  &C_{3pt}(t_f, t, t_i) = \expval{\inter(t_f) J(t) \overline{\inter}(t_i)} - \expval{J(t)}\expval{\inter(t_f) \overline{\inter}(t_i)}, \label{equ:c3}
\end{align}
where $\overline{\mathcal{N}}$~(\(\inter\)) is the nucleon interpolating operator at the source~(sink) timeslice $t_i$~($t_f$), and \(J(t) = \overline{q}\Gamma q\) is the current inserted at time $t$ with spin structure \(\Gamma\).
We note the vacuum subtraction in \cref{equ:c3} which is needed in the case of the scalar charge.

Performing the Wick contractions for the three-point correlation
functions in \cref{equ:c3} leads to quark line connected and
disconnected diagrams.  The connected contributions
are generated through the use of the sequential source method. The
source-sink separation for each three-point function is fixed and
we realise four separations ranging from $t_f-t_i=0.7$~fm up to 1.2~fm.
In order to improve statistics, multiple sources are analysed per
configuration with one, two, three and four measurements being generated for the
four separations ranging from the smallest to the largest.
For the ensembles with periodic boundary conditions in time we make
use of the coherent source technique~\cite{Bratt_2010}.

The disconnected contributions are
constructed by correlating a disconnected loop with a two-point
function.  For the two-point function, typically twenty different
source positions are utilised.  The calculation of the loop is
computationally expensive, as this requires an all-to-all
propagator. To offset this expense we make use of various techniques:
the truncated solver method~\cite{Bali_2010} together with the hopping
parameter expansion~\cite{Thron_1998} and
partitioning~\cite{Bernardson:1993he} in the time direction.  For the
time partitioning we seed the stochastic source on every fourth
timeslice, and then repeat this four times shifting the source by one timeslice
each time. 

The correlation functions are smeared at the source and the sink using
Wuppertal smearing with APE smeared gauge links. The number of
Wuppertal smearing iterations is varied with the pion mass such that
the root-mean-square radius ranges between 0.6~fm and 0.85~fm as the
pion mass decreases from 420~MeV down to the physical point.  For
ensembles with open boundary conditions, the positions of the nucleon
source and sink are chosen such that boundary effects are avoided.

\begin{figure}[ht]
  \centering
  \includegraphics[width=0.75\textwidth]{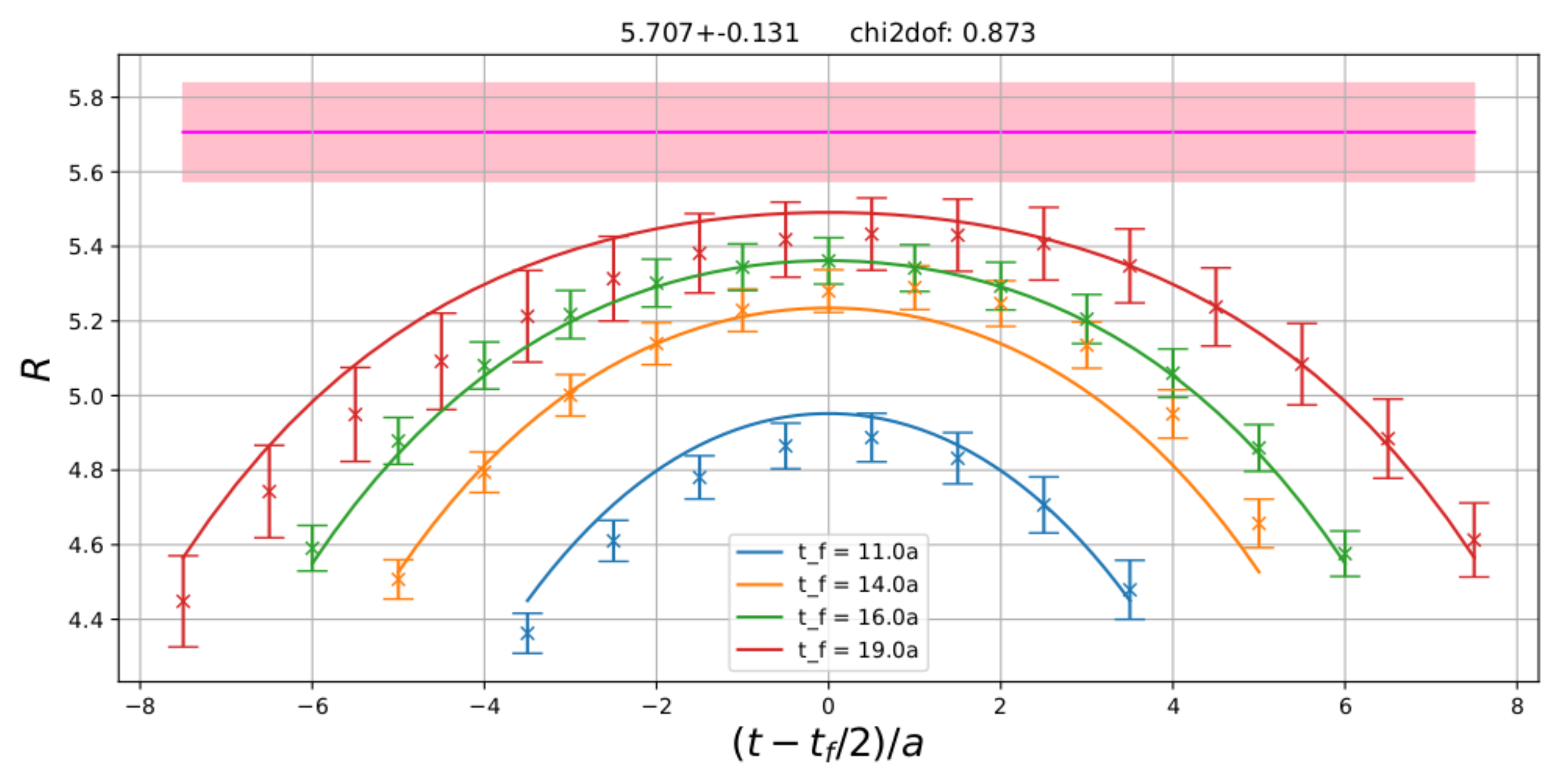}
  \includegraphics[width=0.75\textwidth]{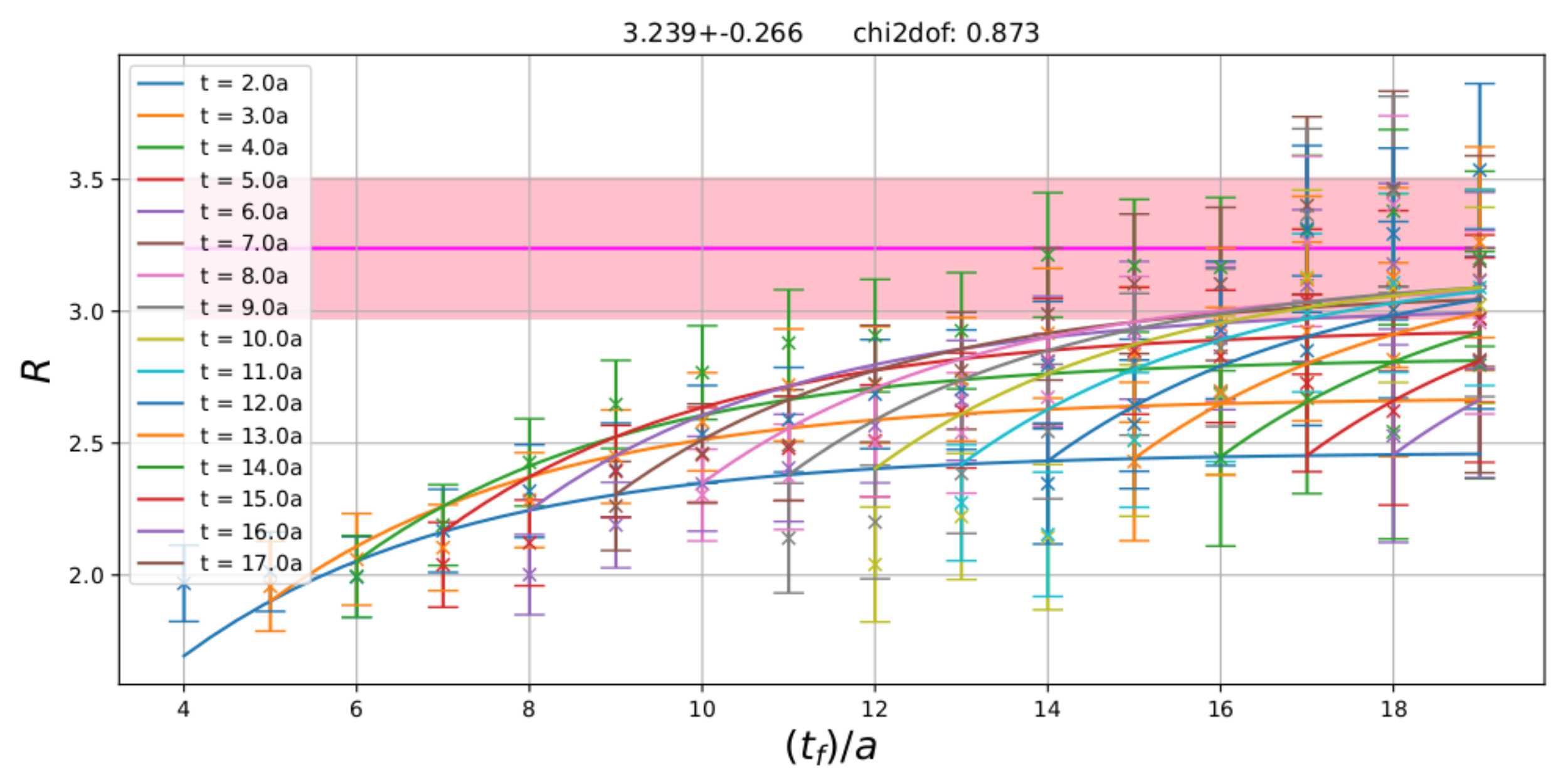}
  \caption{\label{fig:con_plot} (Top) $R(t_f,t)$ for a connected
    three-point function with a scalar current
    $J=\overline{u}\mathds{1}u$ as a function of the current insertion
    time for an ensemble with $m_\pi=345$~MeV, $a=0.0642$~fm, and
    \(L^3\cdot T=3.08^{3}\cdot 8.22\)~fm${}^4$, for four different source-sink
    separations in the range $t_f-t_i=0.7$~fm to $1.2$~fm. (Bottom)
    the ratio for a disconnected three-point function with
    $J=\overline{u}\mathds{1}u$ for multiple current insertion times
    as a function of the sink time. The connected and disconnected
    ratios relevant for extracting scalar, axial, tensor and vector
    charges are fitted simultaneously, with the pink bands indicating
    the ground state matrix elements extracted for the two ratios
    displayed.}
\end{figure}
\subsection{Fitting}\label{sec:fitting}
The spectral decompositions of the two- and three-point correlation
functions, in the limit of large times, read
\begin{equation}
  C_{2pt}(t_f,0) = Z_{1}^2 \me^{-t_f m}\left[ 1 + \frac{Z_{2}^2}{Z_{1}^2} \me^{-\Delta m t_f} \right] + \ldots,
\end{equation}
\begin{equation}
  C_{3pt}(t_f,t,0) = Z_{1}^2 \me^{-t_f m}\left[\expval{J}{1} + \frac{Z_{2} Z_{1}}{Z_{1}^2}\mel{2}{J}{1}\left(\me^{-\Delta m(t_f-t)} + \me^{-\Delta m t}\right) \right] + \ldots,
\end{equation}
where the overlap factors \(Z_{j} \propto \mel{0}{\inter}{j}=Z_{j}^{*}\), and \(\ket{0}\), \(\ket{1}\) and \(\ket{2}\) are the vacuum and the nucleon ground and first excited states, respectively.
The mass gap between the first excited state and the ground state is denoted $\Delta m$.
We fit the ratio of the two- and three-point correlation functions:
\begin{equation}\label{equ:fit form}
  R(t_f,t) = \frac{C_{3pt}(t_f,t,0)}{C_{2pt}(t_f,0)} = \expval{J}{1} + A\mel{2}{J}{1}\left(\me^{-\Delta m(t_f-t)} + \me^{-\Delta m t}\right) + \ldots,
\end{equation}
where the leading constant is the desired matrix element and any
dependence on $t_f$ and $t$ is due to excited state
contamination~(only the leading correction is shown above), which can be
significant.
The excited state spectrum includes multi-particle states as well as
radial excitations. In particular, for ensembles with lighter pion masses,
the \(N(0)\pi(0)\pi(0)\) or \(N(\vec{p})\pi(-\vec{p})\) levels lie below
that of the nucleon's first radial excitation. Furthermore, as the
pion mass approaches the physical point, the excited state spectrum
becomes denser which may lead to difficulties in resolving individual
excited state contributions. 

\begin{figure}[ht]
  \centering
  \includegraphics[width=0.75\textwidth]{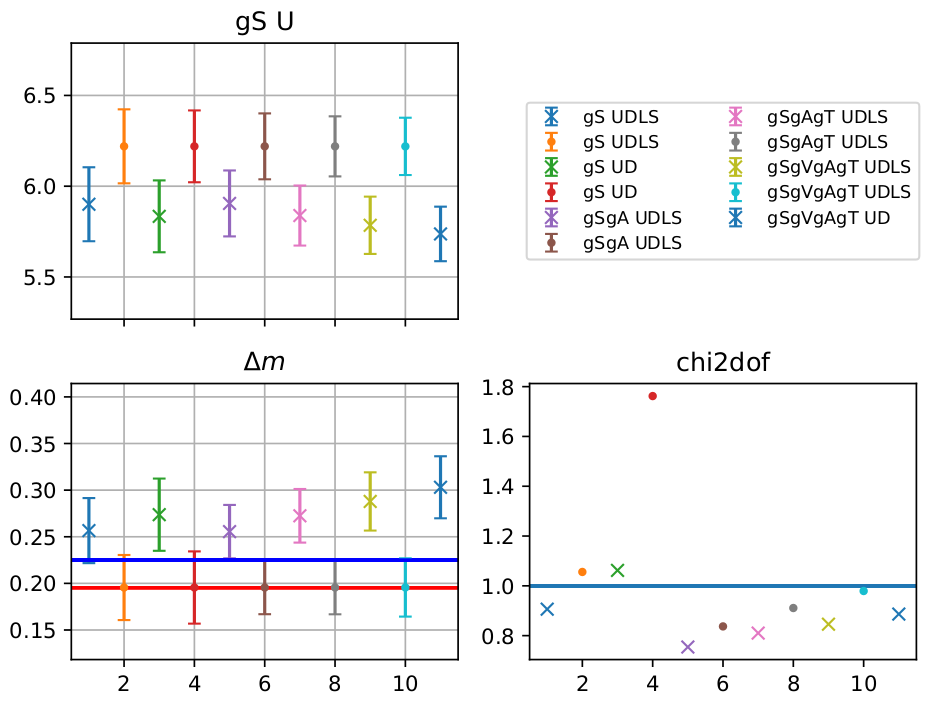}
  \caption{\label{fig:method_var} (Top left) bare connected scalar
    matrix element for $J=\overline{u}\mathds{1}u$ extracted when
    simultaneously fitting to different connected and disconnected
    three-point functions. The legend~(top right) indicates which
    three-point functions (relevant for a particular charge) were
    included, with $U$ and $D$~($L$ and $S$) indicating that
    connected~(disconnected) three-point functions were fitted, with
    $J=\overline{u}\mathds{1}u$ and
    $\overline{d}\mathds{1}d$~($J=\overline{u}\mathds{1}u=\overline{d}\mathds{1}d$
    and $\overline{s}\mathds{1}s$), respectively.  The crosses
    indicate fits where $\Delta m$ is a free parameter, while the
    points correspond to fits where the first excited state mass gap
    is set to the lowest non-interacting $p$-wave \(N\pi\) energy using
    a prior.  (Bottom left) The first excited mass gap extracted. The
    lowest non-interacting $p$-wave \(N\pi\)~($s$-wave
    \(N\pi\pi\)) energy is shown as the red~(blue)
    horizontal line. (Bottom right) The corresponding
    $\chi^2/$d.o.f.\ values.}
\end{figure}

In principle, the mass gap $\Delta m$ can be determined from a fit to
the two-point function, however, as the overlap of the standard
smeared nucleon interpolator with a $N\pi$ or $N\pi\pi$ state is
small, it is often difficult to resolve the lowest excitation.  In
terms of the three-point function, the contribution of this level may
be significant due to an enhanced $\mel{2}{J}{1}$ matrix
element. Alternatively, one can determine the mass gap when fitting to
the ratio $R(t_f,t)$, although, this can be problematic if the excited
state contamination is small.  In order to mitigate these
difficulties, we perform a simultaneous fit to the connected and
disconnected three-point functions corresponding to multiple charges,
enforcing the
same first excited state energy in each case. An example
of such a fit is shown in \cref{fig:con_plot}, where, for brevity,
only the ratios for a scalar current insertion
$J=\overline{u}\mathds{1} u$ are displayed.

To check the assumption that the different charges have the same
dominant excited state, we varied the three-point functions that enter
the fit, as displayed in \cref{fig:method_var}. To further investigate
the sensitivity to the first excited state mass gap, we also performed
fits where $\Delta m$ is set to the lowest non-interacting $p$-wave
\(N\pi\) energy using a prior.  Figure~\ref{fig:method_var} shows that
the matrix element and mass gap are stable as we vary the
three-point functions that are included in the fit. However, there is
a systematic difference between the results with and without the
prior which warrants further study.

\section{Renormalisation}\label{sec:renorm}

Matrix elements determined on the lattice are converted to the
(standard) $\overline{\text{MS}}$ continuum scheme via renormalisation
factors.  When employing Wilson fermions, the flavour singlet and
non-singlet renormalisation factors~($Z^s$ and $Z^{ns}$,
respectively), in general, differ, due to the breaking of chiral
symmetry. This leads to mixing between quark flavours under
renormalisation.  In perturbation theory, the ratio \(r=Z^{s} / Z^{ns} =
1 + \mathcal{O}(\alpha^{n})\), where $r \to 1$ in the
continuum limit~(except in the case of the axial current due to the
anomaly). For the axial and tensor charges, \(n=2\) and \(n=3\),
respectively, suggesting the deviation from one is small. This seems
to be confirmed by non-perturbative determinations of the ratios, see,
e.g.,~\cite{Bali:2016ldn}.

For the scalar, while $Z^{s} / Z^{ns} = 1+\alpha^2$, the ratio is
known to be much larger than one for coarse lattice
spacings. Considering the sigma terms, $\sigma_q=m_{q}g_S^q$, we start with
the renormalisation pattern of
the quark masses~\cite{Bali_2012},
\begin{equation}\label{equ:renorm_matrix}
  \begin{pmatrix}
    m_u(\mu) \\ m_d(\mu) \\ m_s(\mu)
  \end{pmatrix}^{\textrm{ren}}
  = Z^{ns}_m(\mu, a)
  \begin{pmatrix}
    \frac{r_{m} + 2}{3} & \frac{r_{m} - 1}{3} & \frac{r_{m} - 1}{3} \\
    \frac{r_{m} - 1}{3} & \frac{r_{m} + 2}{3} & \frac{r_{m} - 1}{3} \\
    \frac{r_{m} - 1}{3} & \frac{r_{m} - 1}{3} & \frac{r_{m} + 2}{3}
  \end{pmatrix}
  \begin{pmatrix}
    m_u \\ m_d \\ m_s
  \end{pmatrix}^{\textrm{lat}},
\end{equation}
where $r_m=Z^{s}_m / Z^{ns}_m$. 
Defining \(\Tr M = \sum_{q} m_{q}\), \(\Tr g_{S} = \sum_{q}g^q_S\), and \(\widehat{\mathbb{O}}\) to be the renormalised observable \(\mathbb{O}\), we can write
\begin{equation}\label{equ:renorm_m_gs}
  \widehat{m}_{q} = Z_{m} \left( m_{q} + \frac{r_{m}-1}{3}\Tr M \right), \qquad \widehat{g}^q_{S} = Z_{m}^{-1} \left( g^q_{S} + \frac{r_{m}^{-1}-1}{3}\Tr g_{S} \right)
\end{equation}
which gives for the sigma terms
\begin{equation}\label{equ:renorm_sigma}
  \sigma_{q} = \left( m_{q} + \frac{r_{m}-1}{3}\Tr M \right)\left( g^q_S + \frac{r_{m}^{-1}-1}{3}\Tr g_{S} \right).
\end{equation}
Two flavour combinations of note are the pion-nucleon sigma term \(\sigma_{N\pi} = \sigma_{u} + \sigma_{d}\) and the flavour singlet sigma term \(\Tr\sigma = \sum_{q}\sigma_{q}\), which is invariant under renormalisation.

\section{Preliminary Results}\label{sec:res}

\begin{figure}[t]
\begin{minipage}{0.45\textwidth}
  \centering
  \includegraphics[width=\textwidth]{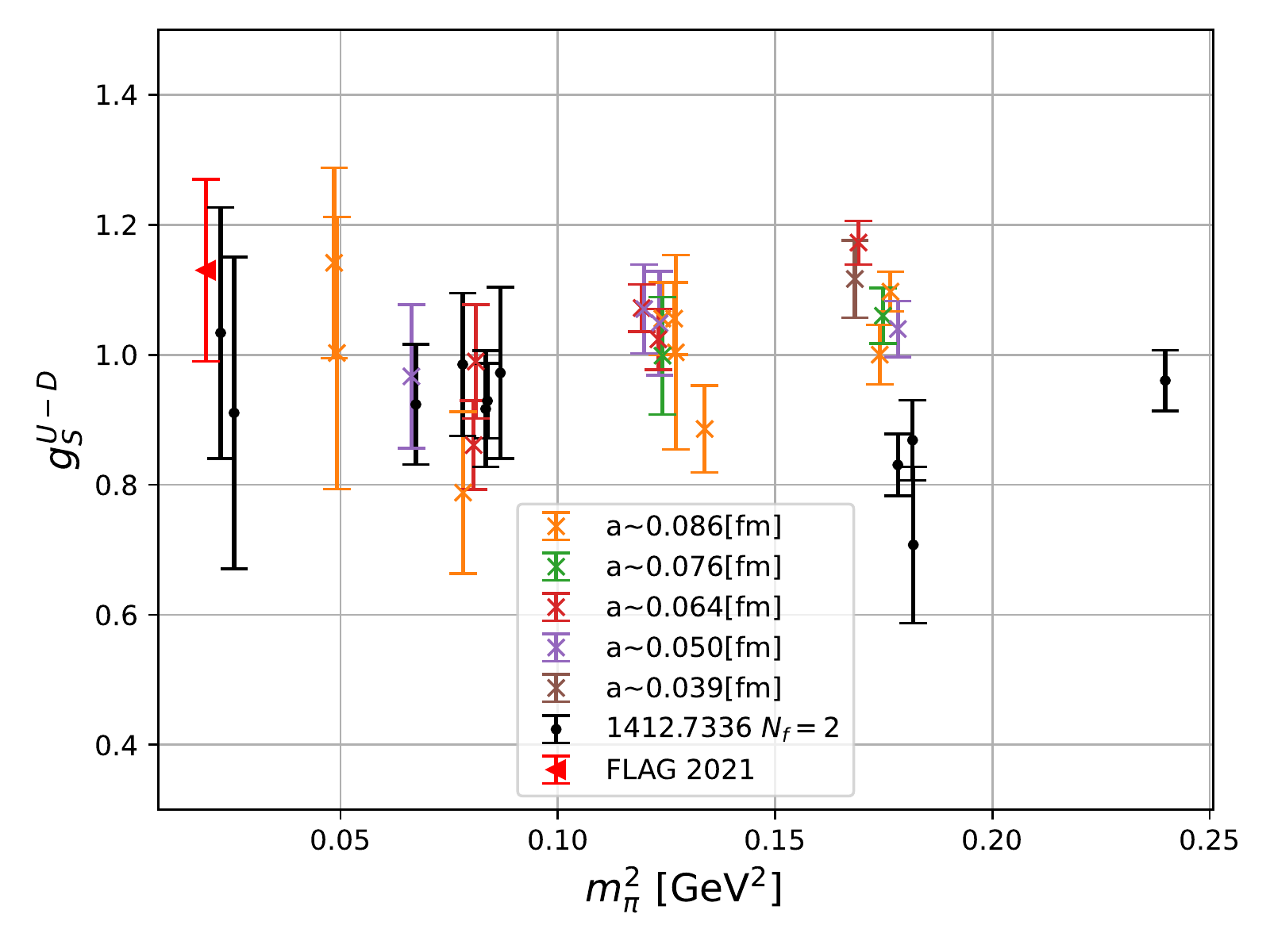}
\end{minipage}
\begin{minipage}{0.45\textwidth}
  \centering
  \includegraphics[width=\textwidth]{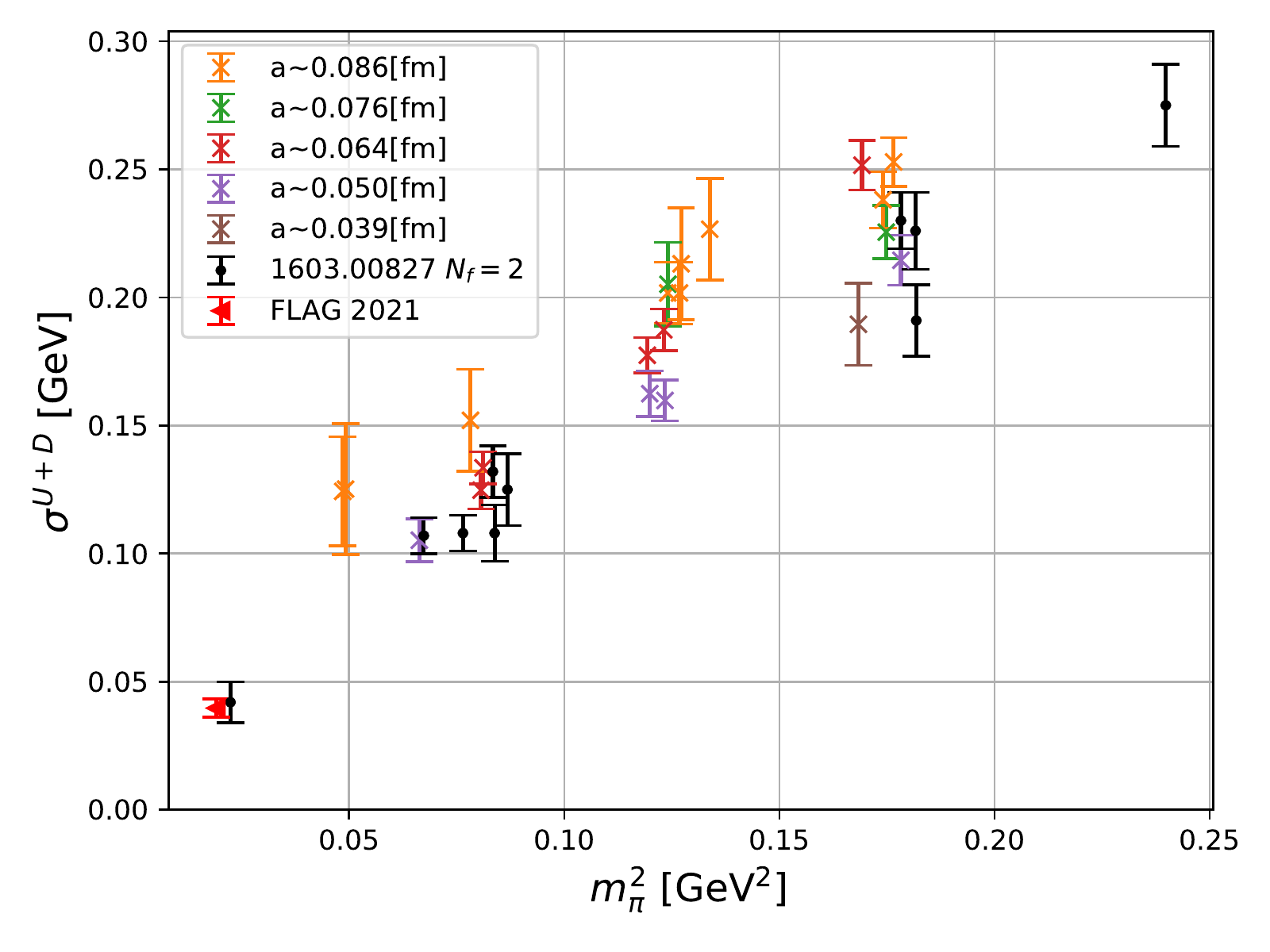}
\end{minipage}
\caption{\label{fig:nf2 comp} Preliminary results for the isovector scalar charge (left) and pion-nucleon sigma terms (right) determined on CLS $N_f=2+1$ ensembles compared to previous RQCD \(N_{f}=2\) results~\cite{Bali:2016lvx} and the recent FLAG average~\cite{FlavourLatticeAveragingGroup:2019iem}.}
\end{figure}
\begin{figure}[h!]
  \centering
  \includegraphics[width=\textwidth]{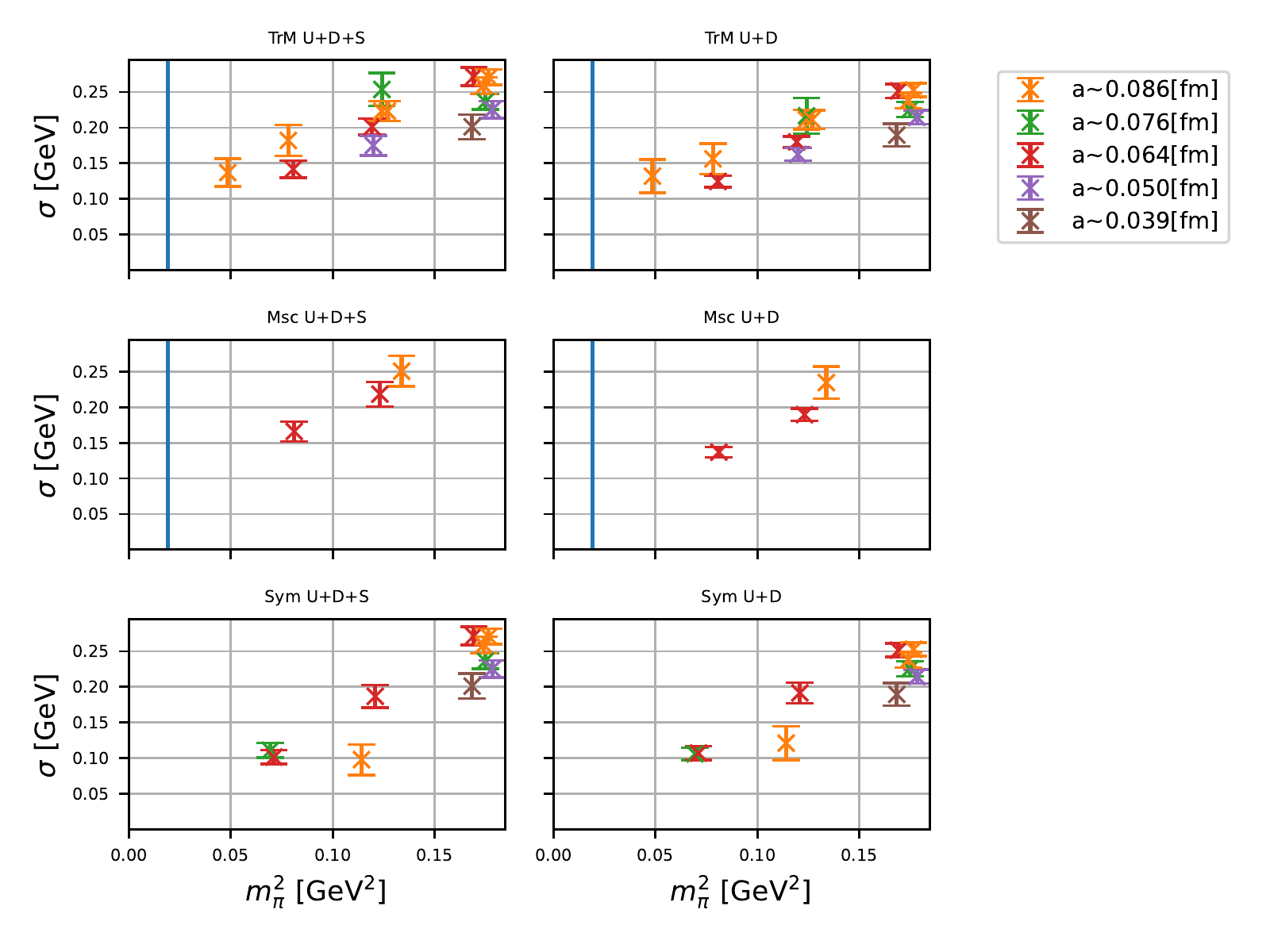}
  \caption{\label{fig:mpi gS} Dependence of the sigma terms on the pion mass squared for the singlet $\sigma_u+\sigma_d+\sigma_s$ flavour combination (left) and the pion-nucleon sigma term (right). The results for the three quark mass trajectories are given separately: (top) the ensembles lie on the trajectory along which the flavour average quark mass is kept constant, (middle) the strange quark mass is approximately constant, (bottom) the light and strange quark masses are equal. Where applicable, the vertical blue line indicates the physical pion mass.}
\end{figure}

\begin{figure}[h!]
  \centering
  \includegraphics[width=\textwidth]{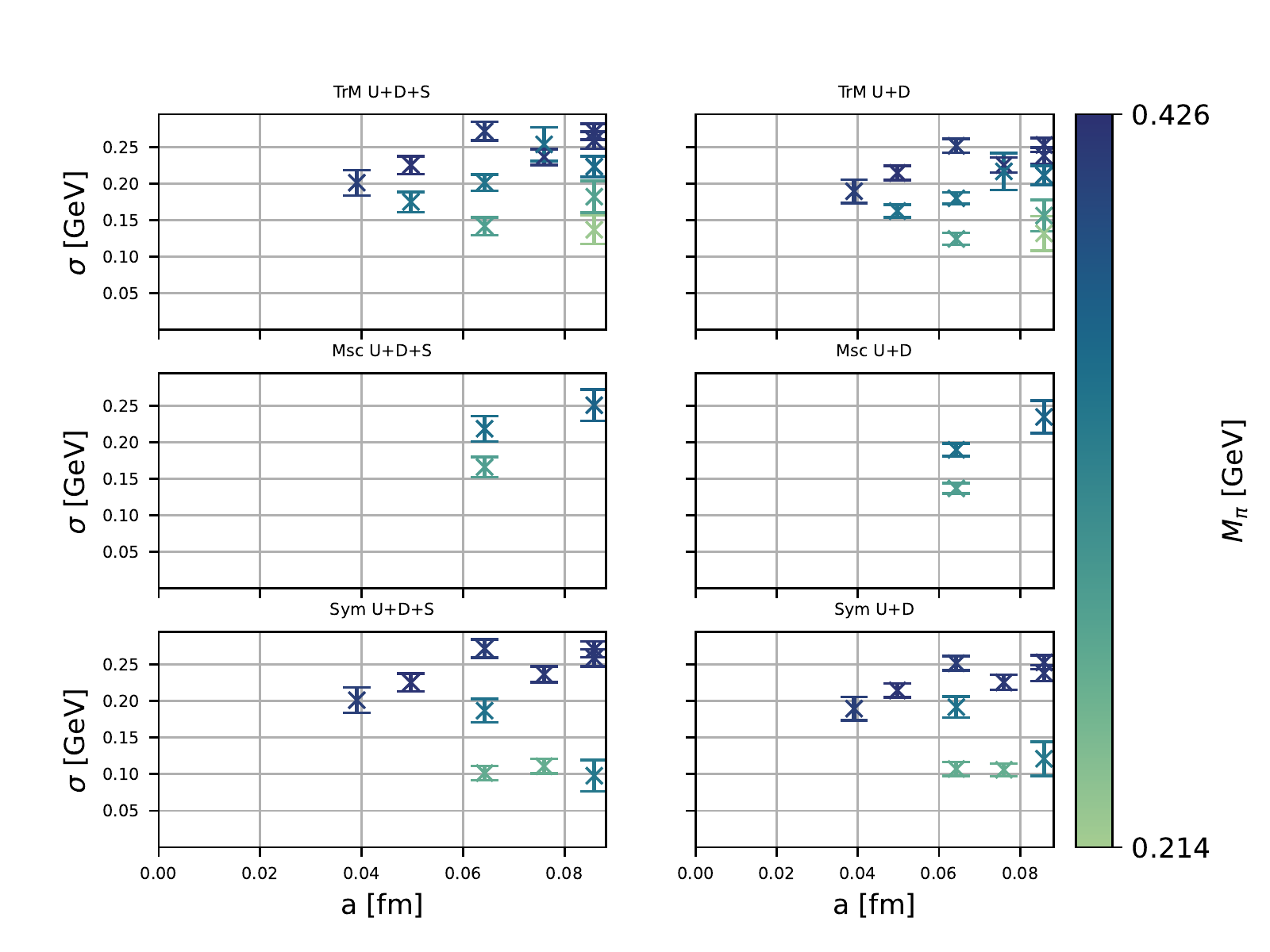}
  \caption{\label{fig:a gS}  Dependence of the sigma terms on the
    lattice spacing for the singlet
    $\sigma_u+\sigma_d+\sigma_s$ flavour combination (left) and 
    the pion-nucleon sigma term (right). The results for the three quark mass
    trajectories are given separately, as in~\cref{fig:mpi gS}. The pion mass in each case is
    indicated by a colour gradient.}
\end{figure}

In \cref{fig:nf2 comp} we show preliminary results for the isovector scalar charge $g_S^{u-d}=g_S^u-g_S^d$ (left) and the pion-nucleon sigma term (right) determined on the $N_f=2+1$ ensembles which lie along the two trajectories that meet at the physical point as a function of the pion mass squared.
Previous \(N_{f}=2\) results from RQCD~\cite{Bali:2016lvx} are also shown for comparison, along with the recent \(N_{f}=2+1\) FLAG average~\cite{aoki20_flag_review}.
Note that results obtained on ensembles which lie along the trajectory, where the flavour average quark mass is kept fixed~(the blue line of \cref{fig:ens}) are only expected to be consistent with the $N_f=2$ results close to the physical point. In \cref{fig:mpi gS,fig:a gS} we present the pion mass and lattice spacing dependence of the results, respectively, for the three different trajectories separately.
The downward trend in the data as the lattice spacing decreases suggests that there may be significant discretisation effects.
In the future we will include further CLS ensembles and perform a combined continuum, quark mass and infinite volume extrapolation.\\

{\bf\noindent Acknowledgments.}
The authors were supported by the European Union’s Horizon 2020 research and innovation programme under the Marie Skłodowska-Curie grant agreement no.\ 813942 (ITN EuroPLEx) and grant agreement no 824093 (STRONG-2020) and by the Deutsche Forschungsgemeinschaft (SFB/TRR-55).
The ensembles were generated as part of the CLS effort using OpenQCD~\cite{Luscher:2012av}, and further analysis was performed using a modified version of CHROMA~\cite{Edwards:2004sx}, the IDFLS solver~\cite{L_scher_2007} and a multigrid solver~\cite{frommer2014adaptive}.
The authors gratefully acknowledge the Gauss Centre for Supercomputing (GCS) for providing computing time through the John von Neumann Institute for Computing (NIC) on JUWELS~\cite{juwels} and on JURECA-Booster~\cite{jureca} at Jülich Supercomputing Centre (JSC).
Part of the analysis was performed on the QPACE~3 system of SFB/TRR-55 and the Athene cluster of the University of Regensburg.

\bibliographystyle{JHEP}
\setlength{\bibsep}{0pt plus 0.3ex}
\bibliography{bib.bib}

\providecommand{\href}[2]{#2}\begingroup\raggedright\begin{thebibliography}{10}

\bibitem{Ji_1995}
X.~Ji et~al., \emph{Qcd analysis of the mass structure of the nucleon},
  \href{https://doi.org/10.1103/physrevlett.74.1071}{\emph{Physical Review
  Letters} {\bfseries 74} (1995) 1071–1074}.

\bibitem{Bruno_2015}
M.~Bruno et~al., \emph{Simulation of qcd with $n_f = 2 + 1$ flavors of
  non-perturbatively improved wilson fermions},
  \href{https://doi.org/10.1007/jhep02(2015)043}{\emph{Journal of High Energy
  Physics} {\bfseries 2015} (2015) }.

\bibitem{Bratt_2010}
J.D.~Bratt et~al., \emph{Nucleon structure from mixed action calculations
  using2+1flavors of asqtad sea and domain wall valence fermions},
  \href{https://doi.org/10.1103/physrevd.82.094502}{\emph{Physical Review D}
  {\bfseries 82} (2010) }.

\bibitem{Bali_2010}
G.S.~Bali et~al., \emph{Effective noise reduction techniques for disconnected
  loops in lattice qcd},
  \href{https://doi.org/10.1016/j.cpc.2010.05.008}{\emph{Computer Physics
  Communications} {\bfseries 181} (2010) 1570–1583}.

\bibitem{Thron_1998}
C.~Thron et~al., \emph{Padé $z_2$ estimator of determinants},
  \href{https://doi.org/10.1103/physrevd.57.1642}{\emph{Physical Review D}
  {\bfseries 57} (1998) 1642–1653}.

\bibitem{Bernardson:1993he}
S.~Bernardson et~al., \emph{{Monte Carlo methods for estimating linear
  combinations of inverse matrix entries in lattice QCD}},
  \href{https://doi.org/10.1016/0010-4655(94)90004-3}{\emph{Comput. Phys.
  Commun.} {\bfseries 78} (1993) 256}.

\bibitem{Bali:2016ldn}
G.S.~Bali et~al., \emph{{Non-perturbative renormalization of flavor singlet
  quark bilinear operators in lattice QCD}},
  \href{https://doi.org/10.22323/1.256.0187}{\emph{PoS} {\bfseries LATTICE2016}
  (2016) 187} [\href{https://arxiv.org/abs/1703.03745}{{\ttfamily
  1703.03745}}].

\bibitem{Bali_2012}
{\scshape QCDSF} collaboration, \emph{{The strange and light quark
  contributions to the nucleon mass from Lattice QCD}},
  \href{https://doi.org/10.1103/PhysRevD.85.054502}{\emph{Phys. Rev. D}
  {\bfseries 85} (2012) 054502}
  [\href{https://arxiv.org/abs/1111.1600}{{\ttfamily 1111.1600}}].

\bibitem{Bali:2016lvx}
{\scshape RQCD} collaboration, \emph{{Direct determinations of the nucleon and
  pion $\sigma$ terms at nearly physical quark masses}},
  \href{https://doi.org/10.1103/PhysRevD.93.094504}{\emph{Phys. Rev. D}
  {\bfseries 93} (2016) 094504}
  [\href{https://arxiv.org/abs/1603.00827}{{\ttfamily 1603.00827}}].

\bibitem{FlavourLatticeAveragingGroup:2019iem}
{\scshape Flavour Lattice Averaging Group} collaboration, \emph{{FLAG Review
  2019: Flavour Lattice Averaging Group (FLAG)}},
  \href{https://doi.org/10.1140/epjc/s10052-019-7354-7}{\emph{Eur. Phys. J. C}
  {\bfseries 80} (2020) 113}
  [\href{https://arxiv.org/abs/1902.08191}{{\ttfamily 1902.08191}}].

\bibitem{aoki20_flag_review}
S.~Aoki et~al., \emph{{FLAG} review 2019},
  \href{https://doi.org/10.1140/epjc/s10052-019-7354-7}{\emph{The European
  Physical Journal C} {\bfseries 80} (2020) 113}.

\bibitem{Luscher:2012av}
M.~L{\"u}scher and S.~Schaefer, \emph{{Lattice QCD with open boundary
  conditions and twisted-mass reweighting}},
  \href{https://doi.org/10.1016/j.cpc.2012.10.003}{\emph{Comput. Phys. Commun.}
  {\bfseries 184} (2013) 519}
  [\href{https://arxiv.org/abs/1206.2809}{{\ttfamily 1206.2809}}].

\bibitem{Edwards:2004sx}
{\scshape SciDAC, LHPC, UKQCD} collaboration, \emph{{The Chroma software system
  for lattice QCD}},
  \href{https://doi.org/10.1016/j.nuclphysbps.2004.11.254}{\emph{Nucl. Phys. B
  Proc. Suppl.} {\bfseries 140} (2005) 832}
  [\href{https://arxiv.org/abs/hep-lat/0409003}{{\ttfamily hep-lat/0409003}}].

\bibitem{L_scher_2007}
M.~Lüscher et~al., \emph{Deflation acceleration of lattice qcd simulations},
  \href{https://doi.org/10.1088/1126-6708/2007/12/011}{\emph{Journal of High
  Energy Physics} {\bfseries 2007} (2007) 011–011}.

\bibitem{frommer2014adaptive}
A.~Frommer et~al., \emph{{Adaptive Aggregation Based Domain Decomposition
  Multigrid for the Lattice Wilson Dirac Operator}},
  \href{https://doi.org/10.1137/130919507}{\emph{SIAM J. Sci. Comput.}
  {\bfseries 36} (2014) A1581}
  [\href{https://arxiv.org/abs/1303.1377}{{\ttfamily 1303.1377}}].

\bibitem{juwels}
{J\"{u}lich Supercomputing Centre}, \emph{{JUWELS: Modular tier-0/1
  supercomputer at the J\"{u}lich Supercomputing Centre}},
  \href{https://doi.org/10.17815/jlsrf-5-171}{\emph{Journal of large-scale
  research facilities} {\bfseries 5} (2018) }.

\bibitem{jureca}
{J\"{u}lich Supercomputing Centre}, \emph{{JURECA: Modular supercomputer at
  J\"{u}lich Supercomputing Centre}},
  \href{https://doi.org/10.17815/jlsrf-4-121-1}{\emph{Journal of large-scale
  research facilities} {\bfseries 4} (2018) }.

\end{thebibliography}\endgroup
\end{document}